\documentclass[aps, prb, twocolumn,amsmath,amssymb,superscriptaddress, nofootinbib,longbibliography,10pt]{revtex4-2}
\usepackage[caption=false]{subfig}
\usepackage{amsfonts,amssymb,amsmath}
\usepackage{graphics}
\usepackage{graphicx}
\usepackage{nicefrac}
\usepackage{cases}
\usepackage{mathrsfs}
\usepackage{enumerate}
\usepackage{mathtools}
\usepackage{textcomp}
\usepackage{verbatim}
\usepackage{bm}          % bold math letters
\usepackage{soul}
\usepackage{cancel}
\usepackage{upgreek}
\usepackage{txfonts}
\usepackage{color}
\usepackage[colorlinks=true, urlcolor=blue, linkcolor=blue]{hyperref}
\usepackage[papersize={8.5in,11in}]{geometry}
\usepackage{esint}
\usepackage[normalem]{ulem}

\renewcommand{\vec}[1]{\boldsymbol{#1}}
\newcommand {\be} {\begin{equation}}
\newcommand {\ee} {\end{equation}}
\newcommand {\e} {\varepsilon}

\usepackage{makecell}
\begin{document}

\title{Disorder-induced liquid-solid phase coexistence in 2D electron systems}

\author{Sandeep Joy}
\affiliation{Department of Physics, Ohio State University, Columbus, OH 43210, USA}
\affiliation{National High Magnetic Field Laboratory, Tallahassee, Florida 32310, USA}
\affiliation{Department of Physics, Florida State University, Tallahassee, Florida 32306, USA}

\author{Brian Skinner}
\affiliation{Department of Physics, Ohio State University, Columbus, OH 43210, USA}

\date{\today}
\begin{abstract}
Recent imaging experiments show a surprisingly robust regime of liquid-solid phase coexistence in a 2D electron system near the quantum melting/freezing transition, with the two phases mixed in mesoscopic domains. Strikingly, the experiments find no noticeable difference in electron density between the liquid and solid domains, which is at odds with both microemulsion scenarios and scenarios in which phase coexistence is driven by fluctuations of a long-ranged disorder potential. Here, we show that such phase coexistence without density difference can be induced by random fluctuations of a short-ranged disorder potential. We further show that disorder tends to stabilize the Wigner Crystal phase to higher densities, which is also consistent with the experiments.

\end{abstract}
%%%%%%%%%%%%%%%%%%%%%%%%%%%%%%%%%%%%%%%%%%%%%%%%%%%%%%%%%%%%%%%%%%%%%%%%
\maketitle
\noindent

\begin{figure}[htb]
\centering
\includegraphics[scale=0.3]{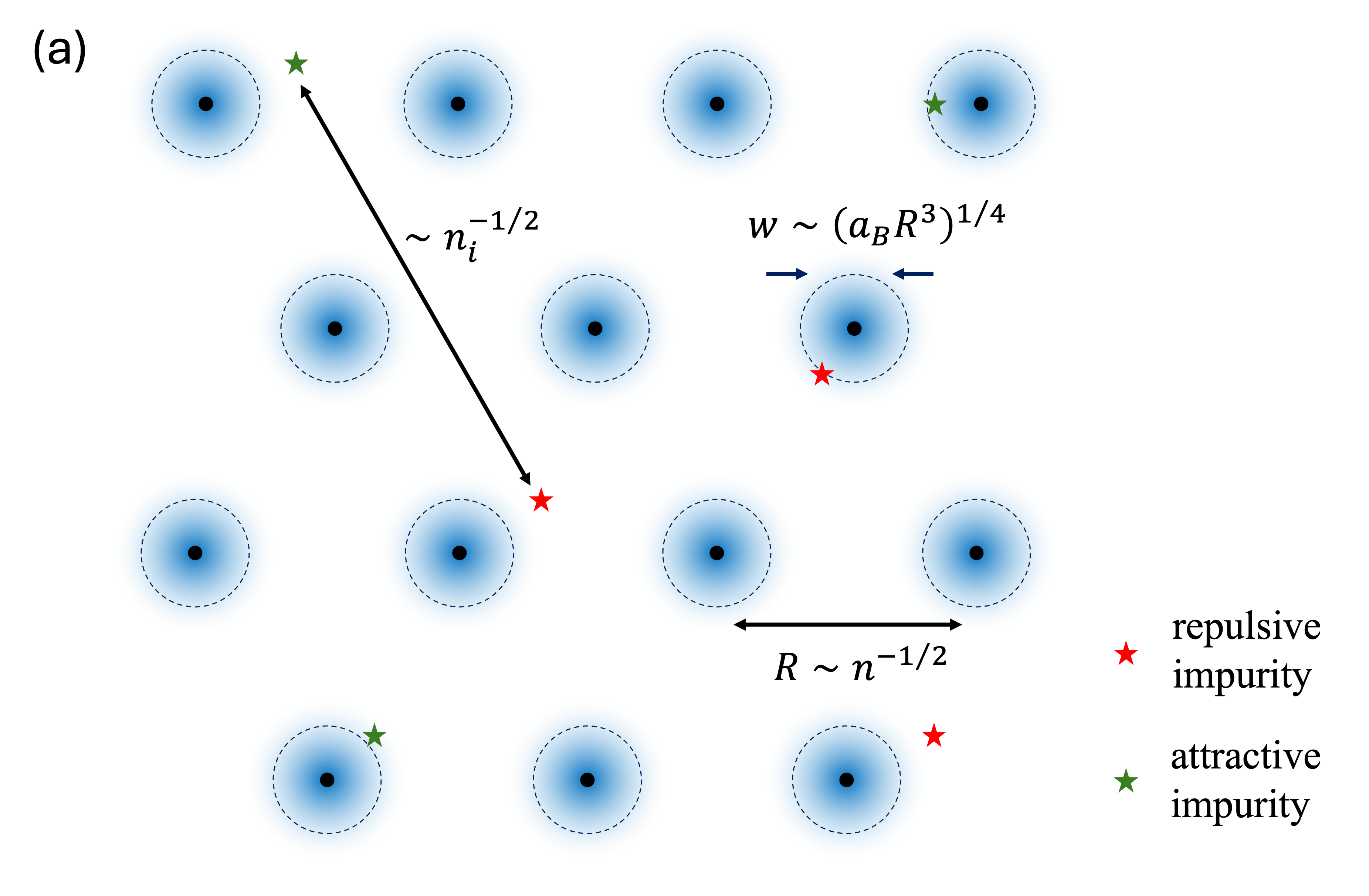}
\includegraphics[scale=0.3]{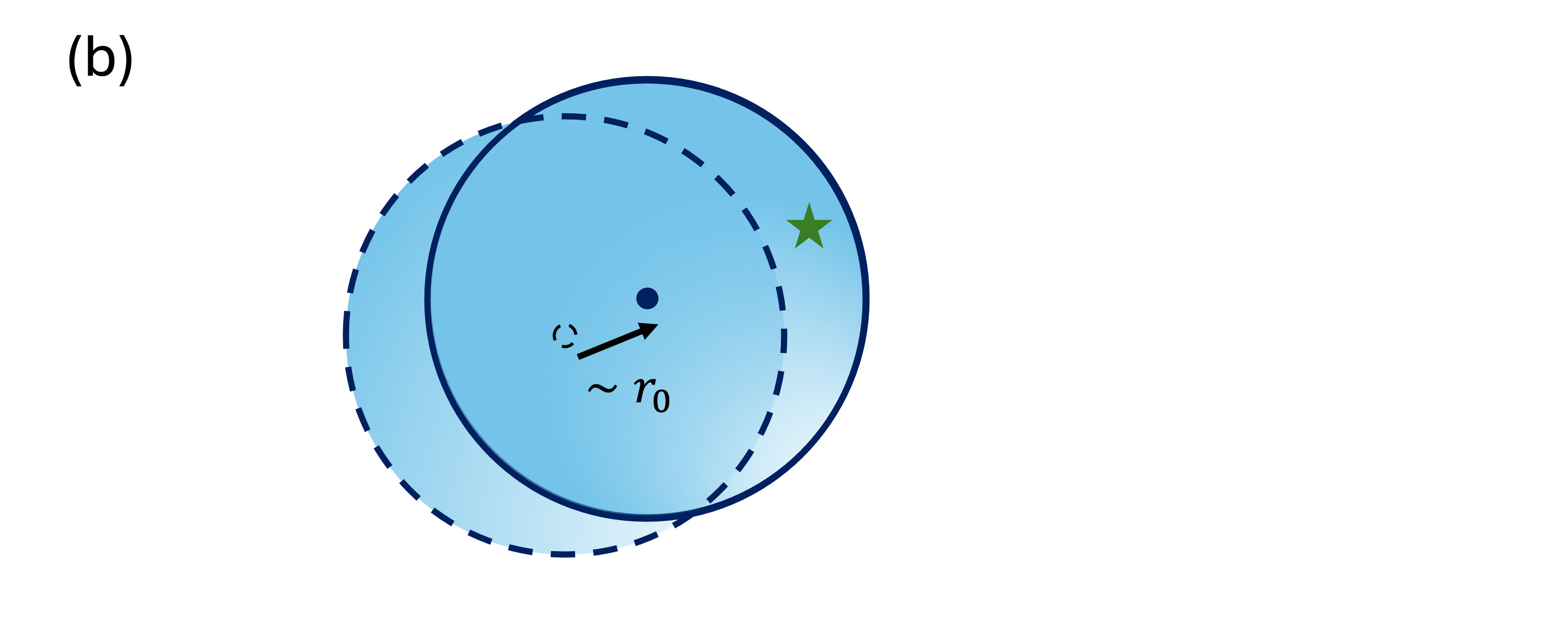}
\caption{(a) A schematic depiction of the WC phase, showing the length scales $R$ (the distance between electrons), $w$ (the size of the electron wavepacket), and $n_i^{-1/2}$ (the distance between impurities). Star symbols denote repulsive (red) and attractive (green) impurities. (b) The wave packet of a particular electron [say, the bottom-left electron in part (a)] displaces by a distance $\sim r_0$ in order to shift into the attractive potential of a nearby impurity.}
\label{fig: undef+def_lattice}
\end{figure}

When the electron density $n$ of an electron system is sufficiently low, the system is unstable (at low temperature) with respect to a phase transition into a solid phase with spontaneously broken translation symmetry. Now, 90 years after Eugene Wigner first predicted \cite{Wigner1934On} the existence of this ``Wigner crystal'' (WC) phase (depicted schematically in Fig.~\ref{fig: undef+def_lattice}), recent experiments \cite{xiang2024quantum, tsui2023direct} have directly imaged the electron solid phase in two-dimensional van der Waals materials. Strikingly, Ref.~\cite{xiang2024quantum} observed a wide regime of electron density for which the WC and liquid phases coexist spatially in mesoscopic domains. Indirect evidence for such coexistence, over a similarly wide window of electron density, was also observed in Ref.~\cite{sung2025anelectronic} using optical spectroscopy. 

%In Ref.~\cite{xiang2024quantum}, scanning tunneling microscopy (STM) was used to directly image both the Wigner crystal and Fermi liquid phase in the bilayer $\text{MoSe}_2$ in the hole doping side. These experiments observed a wide\joy{hole} density regime in which the two phases coexist spatially in mesoscopic domains. In Ref.~\cite{tsui2023direct}, similar STM measurements identified the Wigner solid phase within the lowest Landau level of Bernal bilayer graphene. \sout{though no intermediate phases were observed (the density resolution was significantly smaller than in Ref.~\cite{xiang2024quantum})} In Ref.~\cite{sung2023observation}, cryogenic reflectance and magneto-optical spectroscopy detected signatures of both the solid phase and intermediate phases \joy{in monolayer $\text{MoSe}_2$,} \joy{although it did not involve direct imaging}. 

Theoretically, the nature of the liquid-solid coexistence regime in the 2D electron system (2DES) remains somewhat mysterious. It has been established theoretically for over two decades \cite{spivak2003phase, spivak2004phases, jamei2005universal} that there cannot be a first-order transition directly from a WC phase to a Fermi liquid (FL) phase due to the long-ranged Coulomb interaction between electrons, which effectively prohibits the formation of macroscopic domains with differing charge density. Instead, the authors of Refs.\ \cite{spivak2003phase, spivak2004phases, jamei2005universal} proposed that the transition from WC to FL proceeds through a series of ``microemulsion" phases, in which the two phases are mixed mesoscopically. Finding direct experimental evidence for such microemulsion phases remains a prominent goal of 2D electron physics. A recent work by us \cite{joy2023upper} (following previous work on Coulomb-frustrated
phase separation in Refs.~\cite{lorenzana2001phase1, lorenzana2001phase2, ortix2006frustrated, ortix2007screening, ortix2008coulomb, ortix2009universality}) pointed out that for the clean 2DES the interval $\Delta n$ of average electron density occupied by microemulsions has a strong upper bound that is proportional to the discontinuity in chemical potential $\Delta \mu(n_c)$ between the two phases at the nominally first-order critical density $n_c$. Using Quantum Monte Carlo estimates for $\Delta \mu(n_c)$ \cite{drummond2009phase} yields an apparently very small window of phase coexistence, $(\Delta n)/n_c < 10^{-2}$ \cite{joy2023upper}, which could help to explain why numerical studies have failed to find evidence for a microemulsion regime \cite{drummond2009phase, clark2009hexatic}.

Competing with the microemulsion scenario is a disorder-driven scenario for phase coexistence. In a naive continuum picture, disorder (produced, say, by stray charged impurities embedded in the substrate) produces a continuously varying electrical potential that spatially modulates the electron density by some amount $\delta n_\textrm{dis}$ \cite{shklovskii_completely_1972, ando1982electronic, shklovskii2007simple, adam_self-consistent_2007, falson2022competing}. If one imagines that the electron density fluctuates smoothly over length scales much longer than the lattice constant of the Wigner lattice, then this picture implies a window of phase coexistence $\Delta n = \delta n_\textrm{dis}$, since the electron system locally assumes the WC phase if its local density is $< n_c$ and it assumes the FL phase if the local density is $> n_c$.

The experiments of Ref.~\cite{xiang2024quantum} are especially interesting because they are apparently inconsistent with both scenarios. These experiments directly observe mesoscopic mixing of the liquid and WC phases over a relatively broad window of density, $(\Delta n)/n_c \approx 0.2$, which is in strong contradiction with the upper bound \cite{joy2023upper} on ``intrinsic'' phase coexistence arising from the microemulsion scenario. However, Ref.~\cite{xiang2024quantum} also found that the local electron densities within the WC and FL domains are identical to within $2\%$ at all values of the average electron density, which strongly contradicts the naive disorder-driven scenario.

In this paper, we suggest an explanation for these observations by showing how disorder, even in the form of weak, short-ranged, and spatially uncorrelated impurities, can yield a wide range of phase coexistence that is driven by the structural relaxation of the WC into the disorder potential. Our results also naturally provide an explanation for two other features observed in Ref.~\cite{xiang2024quantum}: (i) the observed critical density $n_c$ is significantly higher than is expected from theoretical calculations with no disorder (i.e., the WC phase is more prominent than expected), and (ii) spatial regions in which the WC phase persists to higher density appear to have an increased concentration of surface impurities. 

Previous theoretical works have considered in detail the effect of impurities on electron solids and charge density waves \cite{fukuyam1978dynamics1, fukuyam1978dynamics2, fukuyama1978pinning, lee1979electric, normand1992pinning, ruzin1992pinning, cha1994orientational, cha1994topological, chitra1998dynamical, fertig1999electromagnetic, fogler2000dynamical, chitra2001pinned, chitra2005zero} (following similar ideas developed in the context of vortex lattices in superconductors \cite{larkin1970effect, larkin1979pinning, blatter1994vortices}). However, these studies primarily focused on how pinning by impurities modifies the static and dynamic responses of a uniform WC or charge density wave phase. Here, we use similar ideas to consider the influence of disorder on phase coexistence near the WC-FL transition. We focus specifically on the weak disorder limit, for which the impurity potential is much weaker than the nearest-neighbor electron-electron interaction, and we show that even in this limit, the disorder can stabilize the WC phase and strongly increase the range of phase coexistence. 
%\joy{It should be pointed out that previous studies have explored the effect of impurities on electron solids and charge density waves  (similarly also in the context of elastic structures formed by vortex lattices in superconductor \cite{larkin1970effect, larkin1979pinning, blatter1994vortices}). However, these studies primarily focused on how impurity-induced pinning influences static properties, such as compressibility, and dynamic properties, such as conductivity response.}

%\section{Perturbation of the WC phase by disorder}

Specifically, we consider a model of disorder in which short-ranged impurities with random sign are distributed randomly in the plane of the electrons with some concentration $n_i$, as depicted in Fig.~\ref{fig: undef+def_lattice}. (While Fig.~\ref{fig: undef+def_lattice} illustrates a situation where $n_i < n$, the theory we develop applies up to much larger impurity densities, as we discuss below.) That is, we consider the impurity potential to consist of random $\delta$-function spikes $\pm \Lambda \delta(\vec{r} - \vec{r_i})$, where $\vec{r}_i$ is the position of an impurity and $\Lambda$ is the strength of the impurity potential. 
%A similar model was considered in Refs.~\cite{fertig1999electromagnetic, fogler2000dynamical, ruzin1992pinning}, although they focused on \brian{****}. \joy{Actually, Ref.~\cite{ruzin1992pinning} considered a much stronger pinning due to donor impurities for which the correction is of the same magnitude as the nearest-neighbor electron-electron interaction. Refs.~\cite{fertig1999electromagnetic} and \cite{fogler2000dynamical} considered a collective pinning approach (analogous to the Larkin-Fukuyama-Lee-Rice model of pinning of an elastic solid by a random potential), which leads to a smaller gain that we will comment on later.}

In the absence of any disorder, and at zero temperature, the WC phase can be described semiclassically as a collection of wave packets of size $w \sim (a_B R^3)^{1/4}$ centered at the lattice sites of a triangular lattice, where $R \sim n^{-1/2} \gtrsim w$ is the lattice constant of the Wigner lattice and $a_B = \hbar^2 /(m e^2)$ is the effective Bohr radius. (Here and throughout this paper, we use CGS units with the dielectric constant $\kappa$ absorbed into $e^2$, so that $e^2/\kappa \rightarrow e^2$.) In this undistorted WC state, the energy due to impurities is zero since the spatial average of the disorder potential is zero. (Introducing an imbalance between $+$ and $-$ impurities only shifts the average value of the electron energy by a constant and does not otherwise alter the arguments we are making in this note.)

However, the electron system can lower its energy by distorting slightly so that electron wave packets shift closer to attractive impurities and farther from repulsive impurities (as depicted in Fig.~\ref{fig: undef+def_lattice}). 
In order to estimate the magnitude of the change in energy resulting from this effect, consider that an electron wave packet centered at position $\vec{r}_0$ (a site of the Wigner lattice) has an electron density
\be 
\rho\left(\vec{r}\right)=\frac{1}{\pi w^{2}}\exp\left[\frac{-\vec{r}^{2}}{w^{2}}\right],
\ee 
and the potential energy felt by that electron due to an impurity at position $\vec{r}_i$ is 
\be 
U_\textrm{imp} = \pm \Lambda \rho(\vec{r}_i).
\ee 
The corresponding force on the wave packet due to the impurity is $\vec{F}_i(\vec{r}_i) = \mp \Lambda \left. \left( \nabla \rho(\vec{r})\right) \right|_{\vec{r} = \vec{r}_i}$. When averaging over all possible impurity positions and signs, the expectation value of the force on a given electron is zero. But the mean-square force $(\Delta \vec{F})^2$ is nonzero, and can be calculated as
\be 
(\Delta \vec{F})^2 = \int d^2 \vec{r}_i \, n_i \, \left[ \vec{F}_i(\vec{r_i}) \right]^2 ,
\label{eq: sumsq}
\ee 
where the quantity $d^2 \vec{r}_i n_i$ represents the probability that an impurity is located within an infinitesimal area $d^2 \vec{r}_i$ of the location $\vec{r}_i$, so that Eq.~(\ref{eq: sumsq}) represents a simple sum of variances in the net force produced by each impurity. Evaluating this integral gives a root-mean-square force 
\be 
\sqrt{(\Delta \vec{F})^2} = \sqrt{n_i \Lambda^2 / \pi w^4}.
\ee
This quantity represents the typical force experienced by an electron in its undisplaced Wigner lattice site. Countering this force is the elastic force $- m \omega^2 \vec{r}_0$ associated with a small displacement $\vec{r}_0$ of the wave packet from the Wigner lattice site, where $m \omega^2 = \gamma e^2 n^{3/2}$, with $\gamma \approx 4.44$ a numerical constant \cite{ruzin1992pinning, joy2022wigner}. 
Assuming that $|\vec{r}_0| \ll w$, one can equate these two forces to arrive at a typical equilibrium displacement $r_0 = \sqrt{(\Delta \vec{F})^2} / m \omega^2 \sim \sqrt{n_i \Lambda^2 m / \hbar^2}$, and therefore a reduction in energy $\e_\textrm{imp} = -\frac12 m \omega^2 r_0^2$, or
\be 
\e_\textrm{imp} = - \frac{n_i}{2 \pi} \frac{\Lambda^2 m}{\hbar^2}.
\label{eq: eimp}
\ee
Here, we have used the equality $w = \sqrt{\hbar/(m \omega)}$. Notice that our assumption $r_0 \ll w$ is equivalent to $\e_\textrm{imp} \ll \hbar \omega$, or in other words that the correction due to disorder is small compared to the first quantum correction to the energy of the Wigner crystal. In this limit one can indeed consider that electron wave packets are shifting rigidly in the disorder potential and are not distorted significantly.

%A single attractive impurity within the interior of the wave packet lowers the energy of the electron by an amount $\sim - \Lambda / w^2$, while a displacement of the wave packet by an amount $r$ from the Wigner lattice site increases the energy by an amount $\sim e^2 r^2/R^3$ \cite{ruzin1992pinning}. Equating these two energy scales implies that displacements are only favorable if the impurity resides within a distance $\sim r_0$ of the perimeter of the wave packet, where $r_0 \sim \sqrt{\Lambda R^3/(e^2 w^2)}$. Throughout this note, we assume that the impurity potential is weak enough that the impurity energy $\sim \Lambda /w^2$ is much smaller than the first quantum correction to the electron energy, $\sim \sqrt{e^4 a_B / R^3}$, so that the size and shape of the wave packet are mostly undistorted by the impurity potential. (Substituting the value of $w$ gives a condition $\Lambda \ll e^2 a_B$.)

One can also consider the contribution to $\e_\textrm{imp}$ arising from collective transverse shear of the WC (as discussed, for example, in Refs.\ \cite{larkin1979pinning, fukuyam1978dynamics1, fukuyam1978dynamics2, fukuyama1978pinning, lee1979electric, ruzin1992pinning, fertig1999electromagnetic, fogler2000dynamical}), rather than from displacement of individual electron wave packets. 
This collective relaxation process ultimately destroys the long-range order of the Wigner lattice and determines the finite correlation length of the WC.  However, we show in Appendix~\ref{sec: shear} that it produces a contribution to the electron energy that is parametrically smaller than Eq.~(\ref{eq: eimp}) when the impurity positions have a correlation length longer than $w$, and in the limit where there is no correlation whatsoever between impurity positions it produces a contribution that is the same order of magnitude as $\e_\textrm{imp}$.

% We also note that the result in Eq.~(\ref{eq: eimp}) is different and larger in magnitude than the result that arises from considering relaxation of the WC state through collective transverse shear (as considered, for example, in ).  We discuss this collective relaxation through transverse shear in Appendix~\ref{sec: shear}.  

The disorder-induced correction to the energy of the FL phase can similarly be estimated using the following semiquantitative argument. In the liquid phase, an impurity induces a change in electron density $\delta n$ over some region of size $r$. If we suppose, for concreteness, that the impurity is attractive, then $\delta n$ is positive, and the potential energy is lowered by an amount $- \Lambda \delta n$ due to the perturbation. This gain should be balanced against both an increase in kinetic energy $(\delta n)^2 r^2 / \nu$, where $\nu \sim \hbar^2/m$ is the density of states and a Coulomb self-energy $\sim (e \delta n r^2)^2/r$. Minimizing the total energy with respect to $\delta n$ gives an energy $E_i \sim -\Lambda^2/[r^2(\nu^{-1} + e^2 r)]$ associated with a single impurity. This energy is minimized when $r$ corresponds to the smallest scale over which the Fermi liquid can produce a perturbation in density, or in other words, to $r \sim n^{-1/2} = R$, the inverse Fermi wavelength. Inserting this value gives $E_i \sim - \Lambda^2 n m / [\hbar^2 (1 + r_s)]$ (where $r_s$ is the dimensionless interaction parameter, $r_s\sim R/a_B$). The corresponding energy per electron is $\e_\textrm{imp}^\textrm{FL} = E_i n_i/n \sim -\Lambda^2 n_i m / [\hbar^2(1 + r_s)]$. Notice that this value is smaller than the estimation in Eq.~(\ref{eq: eimp}) by a factor $1/(1 + r_s)$. Since $r_s \gtrsim 30$ near the transition \cite{drummond2009phase}, we conclude that the FL phase is less able to accommodate disorder energetically than the WC phase. Of course, this simple description of the response of the FL phase amounts to making an assumption of linear screening, which is, in general, not valid very close to the FL-WC transition, where the effective mass is apparently strongly renormalized by interactions and the Fermi energy $E_F$ becomes very small \cite{falson2022competing}. But in general, our description can be applied so long as the impurity potential $\Lambda$ is small enough that $E_i \ll E_F$, so that our conclusion of a weaker correction to energy per electron in the FL phase holds for sufficiently weak impurity potentials..

\begin{figure}[htb]
\centering
\includegraphics[scale=0.5]{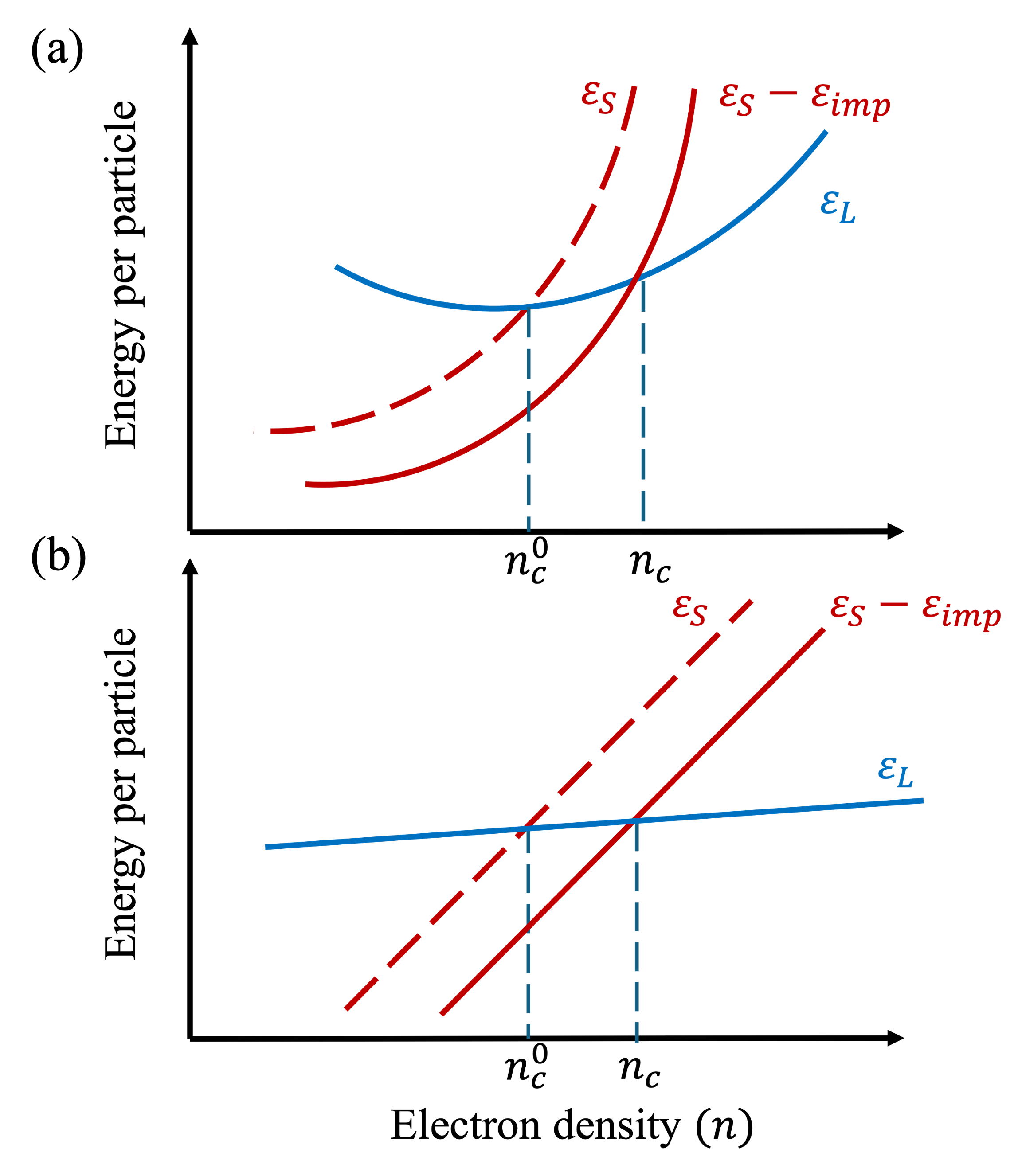}
\caption{Schematic plots of the energy per electron as a function of electron density. The red lines correspond to the energy per electron of the WC phase without disorder ($\e_S$, dashed) and with disorder $\varepsilon_S - \varepsilon_{imp}$, solid). The blue solid line depicts the energy per electron of the liquid phase, which is not significantly altered by the disorder. Thus, the disorder shifts the critical density associated with the phase transition from $n_c^0$ to a higher value $n_c$. Figures (a) and (b) plot the same quantities, with (b) corresponding to a linearized (zoomed-in) version of the plot in (a).}
\label{fig: energy_per_electron}
\end{figure}

This argument allows us to conclude that the WC phase is effectively \emph{stabilized} by the disorder, and the critical concentration $n_c$ associated with the WC-FL phase transition is shifted toward larger values, as apparently observed in Ref.~\cite{xiang2024quantum}. This shift is depicted schematically in Fig.~\ref{fig: energy_per_electron}.

Linearizing the relation $\e(n)$ around $n_c$ for each of the two phases, we find that for weak impurities the critical density associated with the phase transition is increased such that
\be 
n_c = n_c^0 \left( 1 + \frac{\left| \e_{imp} \right|}{\Delta \mu(n_c^0)} \right),
\label{eq: ncshift}
\ee 
where $n_c^0$ is the critical density in the absence of disorder. Notice that Eq.~(\ref{eq: ncshift}) implies that samples with higher impurity concentration exhibit a WC phase that persists to higher electron density $n$. 
%\joy{A smaller $\Delta \mu$ (found in QMC calculations \cite{drummond2009phase}) is conducive to stabilizing the WC phase to higher concentrations.} 
This conclusion is consistent with observations \cite{privatecommunication} by the authors of Ref.~\cite{xiang2024quantum}.

%\section{Phase coexistence due to fluctuations in impurity concentration}

We now discuss how our result for the correction in energy to the WC phase implies a regime of phase coexistence. As derived in Eq.~(\ref{eq: eimp}), the average energy per electron within the WC phase is reduced by an amount that depends linearly on the impurity concentration. However, the impurity concentration within a particular spatial region of size $L$ is subject to statistical fluctuations. Specifically, the number of impurities $N_i^L$ inside a region of size $L \gg n_i^{-1/2}$ is $N_i^L \sim n_i L^2 \pm \sqrt{n_i L^2}$ (assuming that impurity positions are uncorrelated), so that the impurity concentration within the region fluctuates from one $L$-sized region to another by an amount  
\be 
\delta n_i^L \sim \frac{\sqrt{n_i}}{L}.
\ee
This spatial fluctuation in the local impurity concentration implies that the energy per electron also fluctuates as one considers different spatial regions over which to average. Specifically, using Eq.~(\ref{eq: eimp}), the fluctuation in the energy per electron from one $L$-sized region to another is
\be 
\delta \e_{imp}^L \sim \delta n_i^L \frac{\Lambda^2 m}{\hbar^2}.
\label{eq: deltaeimpL}
\ee 

Let us now consider whether such spatial fluctuations can produce significant spatial coexistence between FL and WC phases in the vicinity of the critical point $n_c$. We assume everywhere that the FL and WC phases have equal electron concentration $n$, as observed in the experiment \cite{xiang2024quantum}, so that there is no energy penalty associated with long-ranged Coulomb interactions.

Suppose first that the average electron density $n$ is exactly equal to the critical density $n_c$ at which the two phases have equal energy (including the downward shift $\e_{imp}$ of the WC phase arising from the impurity potential). Consider now the process of dividing the system up into large $L$-sized regions. If one ignores the effects of surface tension, then those regions which by chance have higher-than-average impurity concentration will assume the WC phase (since the downward energy correction $\e_{imp}$ is proportional to $n_i$), while those regions with lower-than-average impurity concentration will assume the FL phase. In this way, each $L$-sized region is able to lower its energy by an amount $\sim \delta \e_{imp}^L \cdot (L/R)^2$, or in other words the change in energy per electron is $\sim -\e_{imp}^L \propto 1/L$ relative to the uniform phase. The impurity energy, therefore, favors the system forming small-sized phase domains in order to maximally take advantage of statistical fluctuations in impurity concentration.

Competing with this tendency to form small domains is the surface tension between the two phases, which arises from microscopic interactions between electrons at the boundary between the two phases. This surface tension implies an energy cost $\gamma L$ per length $L$ of the perimeter, where $\gamma$ is the surface tension, and therefore adds an energy cost $\sim \gamma / L$ per unit area or $\sim \gamma R^2/L$ per electron. The surface tension, therefore, favors large domains, which reduce the total interfacial energy of the system. 

Notice that both the reduction in energy due to impurities and the increase in energy due to surface tension both scale with the same power of the domain size $L$. Consequently, there is no optimal domain size $L$ within this description, and instead, the value of $L$ becomes as small as possible if
\be 
\gamma \lesssim n \sqrt{n_i} \frac{\Lambda ^2 m}{\hbar^2}.
\label{eq: gamma inequality}
\ee 
% \be 
% \gamma \lesssim e^2 n \sqrt{n_i a_B^2} \left( \frac{\Lambda}{e^2 a_B} \right)^{2}.
% \label{eq: gamma inequality}
% \ee 
When this inequality is violated, the value of $L$ instead becomes as large as possible, or in other words, the system assumes a uniform phase without domain walls.

Let us assume without proof that Eq.~(\ref{eq: gamma inequality}) is satisfied. (Such a situation is plausible, given the very small difference in chemical potential $\Delta \mu(n_c)$ between the WC and FL phases observed numerically, which presumably implies a small surface tension $\gamma$ \cite{drummond2009phase}).  %\brian{***** the following paragraph needs to be fixed. eqs 10 and 11 are wrong ****} 
In this case, the system minimizes its energy by making $L$ so small that only an order-1 number of electrons are significantly displaced by impurities within the domain. Since, as described above, the probability that any given electron will have an impurity within the area occupied by its wave packet is $n_i w^2$, we arrive at the conclusion that $(n_i w^2)(L/R)^2 \sim 1$, or in other words that the domain size is
\be 
L = L_\text{min} \sim \frac{R}{\sqrt{n_i w^2}}.
\ee 
This expression assumes a sufficiently sparse impurity concentration $n_i w^2 \ll 1$, or $n_i a_B^2 \ll r_s^{-1/2}$. Below, we briefly discuss the opposite limit of large impurity concentration $n_i w^2 \gg 1$ (many impurities within each electron wave packet).

\begin{figure}[tb]
\centering
\includegraphics[scale=0.46]{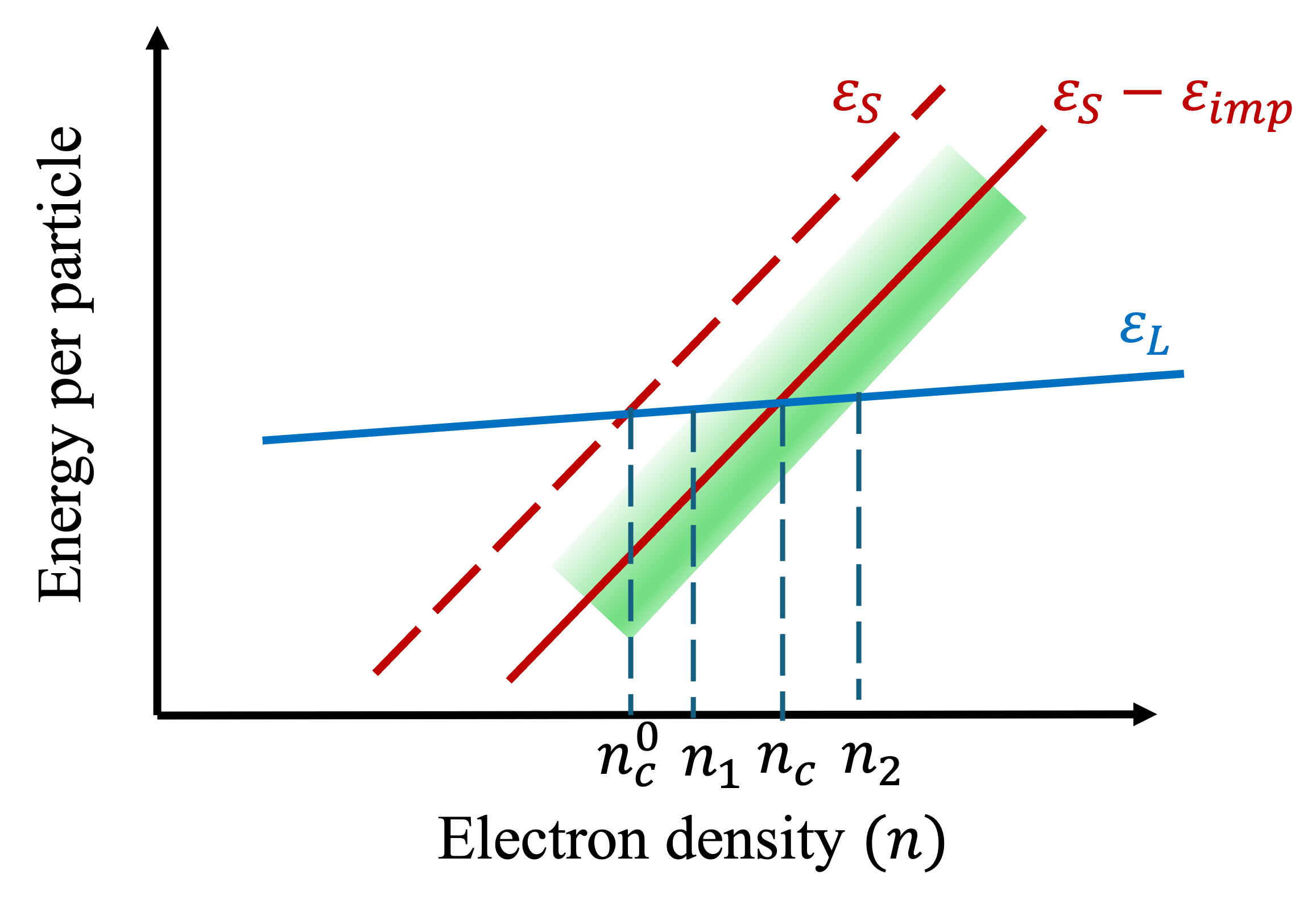}
\caption{Due to the spatial fluctuations of the impurity density, the energy per electron of the WC phase is smeared out by $\pm \delta \varepsilon_{imp}^{L_\text{min}}$. This smearing implies a spatially varying melting density, leading to phase coexistence in the interval $n_1 > n > n_2$. In the figure, the height of the green rectangle represents the quantity $\delta \e_{imp}^{L_\text{min}}$.}
\label{fig: phase_coext}
\end{figure}

Inserting the value $L = L_\text{min}$ into Eq.~(\ref{eq: deltaeimpL}), we arrive at the conclusion that the energy per electron has spatial fluctuations
% \be 
% \delta \e_{imp}^{L_\text{min}} \sim e^2 n_i a_B \, (n a_B^2)^{-1/4} \left( \frac{\Lambda}{e^2 a_B} \right)^{7/4}.
% \ee 
\be 
\delta \e_{imp}^{L_\text{min}} \sim  \frac{\Lambda^2 m }{\hbar^2} \frac{n_i w}{R}.
\ee 
from one domain to another. One should, therefore, think that the energy of the WC phase also has spatial fluctuations of the same magnitude, as depicted in Fig.~\ref{fig: phase_coext}. Consequently, there is a finite window of average density over which the two phases can coexist spatially, which corresponds to the range of density for which the energy per electron of the two uniform phases are within $\delta \e_{imp}^{L_\text{min}}$ of each other.

Assuming that this density range $\Delta n \ll n_c$, we arrive at the conclusion
\be 
\frac{\Delta n}{n_c} \sim \frac{\delta \e_{imp}^{L_\text{min}}(n_c)}{\Delta \mu(n_c)} \sim \frac{\Lambda^2 m }{\hbar^2 \Delta \mu} n_i r_s^{-1/4}.
\ee 
%where, as before, $n_c \approx n_c^0$.
Notice that the width $\Delta n$ of the phase coexistence regime is large when $\Delta \mu$ is small.
%\joy{(similar to the case of Eq.~\ref{eq: ncshift}, where small  $\Delta \mu$ lead to higher melting density)}. 
In other words, this mechanism of disorder-induced phase coexistence is most prominent precisely under conditions where the disorder-free microemulsion scenario is negligible.

In the limit of sufficiently large impurity concentration that $n_i w^2 \gg 1$ [but still $n_i w^2 \ll (e^2 a_B/\Lambda)^2$, so that we can assume linear elastic response of the Wigner lattice], the optimal domain size $L_\text{min}$ at the transition becomes comparable to the Wigner lattice constant, $L_\text{min} \sim R$. In this limit, the width of the phase coexistence regime becomes
\be 
\frac{\Delta n}{n_c} \sim \frac{\Lambda^2 m}{\hbar^2 \Delta \mu} \frac{\sqrt{n_i}}{a_B} r_s^{-1}.
\nonumber 
\ee

In summary, we have shown that short-ranged disorder can stabilize the WC phase and lead to a significant regime of electron density in which the FL and WC phases are mixed mesoscopically. While such mesoscopic mixing can resemble microemulsion physics qualitatively, in our case, it is driven by disorder and can be distinguished visually by a qualitatively different dependence of the domain size on the electron density and by the tendency of WC domains to occupy regions of the sample with larger impurity concentration. Importantly, the scenario we discuss does not involve any difference in local electron density between liquid and solid domains, which is a crucial ingredient in previous disorder-driven explanations for phase coexistence \cite{shklovskii_completely_1972, ando1982electronic, shklovskii2007simple, adam_self-consistent_2007, falson2022competing, joy2023upper}. 

Of course, in real experimental samples, disorder arises from both short-ranged impurities (e.g., vacancies or isovalent substitutions in the crystal) and charged impurities (dopant impurities with different valence); our model realistically applies only to uncharged impurities and does not capture more subtle effects associated with long-ranged fluctuations of the Coulomb potential that are produced by charged impurities. Nonetheless, in the experiments of Ref.~\cite{xiang2024quantum, privatecommunication}, uncharged impurities are much more prevalent on the surface of the samples than charged impurities, so it is reasonable to think that the effect we discuss here is the dominant driver of phase coexistence. 

Finally, while we have focused primarily on the STM experiments of Ref.~\cite{xiang2024quantum}, we note that there are a range of experiments that find evidence for smearing of the phase transition from WC to FL (e.g., Refs.~\cite{yoon1999wigner, knighton2018evidence, li2019evidence, hossain2020observation, falson2022competing, yang2023cascade, sung2025anelectronic}). Whether our results apply to some or all of these experiments remains to be considered.

\ 

\textit{\textbf{Acknowledgments:}} The authors are grateful to Mike Crommie, H. A. Fertig, Zhehao Ge, Sarang Gopalakrishnan, Zehao He, and B.~I.~Shklovskii for useful discussions. S.J. acknowledges the support from Florida
State University through the FSU Quantum Postdoctoral
Fellowship. This work was supported by the NSF under Grant No.~DMR-2045742.

%%%%%%%%%%%%%%%%%%%%%%%%%%%%%%%%%%%%%%%%%%%%%%%%%%%%%%%%%%%%%%%%%%%%%%%%%%%%%%%%%%%%%%%%%%

%%%%%%%%%%%%%%%%%%%%%%%%%%%%%%%%%%%%%%%%%%%%%%%%%%%%%%%%%
\bibliography{ref}
%%%%%%%%%%%%%%%%%%%%%%%%%%%%%%%%%%%%%%%%%%%%%%%%%%%%%%%%%
\widetext
% \clearpage
% \begin{center}
% \textbf{\large Supplementary Information for ``Disorder-induced liquid-solid phase coexistence in 2D electron systems"}
% \end{center}
% %%%%%%%%%% Merge with supplemental materials %%%%%%%%%%
% %%%%%%%%%% Prefix a "S" to all equations, figures, tables and reset the counter %%%%%%%%%%
% \setcounter{equation}{0}
% \setcounter{figure}{0}
% \setcounter{table}{0}
% \setcounter{page}{1}
% \makeatletter
% \renewcommand{\theequation}{S\arabic{equation}}
% \renewcommand{\thefigure}{S\arabic{figure}}
% \renewcommand{\bibnumfmt}[1]{[S#1]}
% \renewcommand{\citenumfont}[1]{S#1}
% %%%%%%%%%% Prefix a "S" to all equations, figures, tables and reset the counter %%%%%%%%%%

% \begin{center}
% Sandeep Joy, Brian Skinner \\
% \textit{Department of Physics, Ohio State University, Columbus, OH 43210, USA} \\
% (Dated: \today)
% \end{center}

\appendix
%\section{S1. Collective pinning of the Wigner Crystal due to impurities}
\section{Relaxation of the Wigner Crystal due to collective shear motion}
\label{sec: shear}

In the main text, we considered the process of the WC state relaxing due to small displacements of individual electrons in the Wigner lattice, and we found that the energy of the WC state is lowered by an amount proportional to $n_i \Lambda^{2}$ [Eq.~(\ref{eq: eimp})]. In this appendix, we consider an alternative mechanism by which the Wigner lattice can respond to disorder, which is through collective shearing. Such shearing motion has been considered extensively before; the arguments we present below are largely adapted from those in Refs.~\cite{ruzin1992pinning, fertig1999electromagnetic}. We show that, for the model under consideration, the reduction in energy due to shearing is no larger than Eq.~(\ref{eq: eimp}), and when the impurity positions have significant spatial correlations it can be parametrically smaller.
%We show here that for the model we are considering, the reduction in energy due to shearing is much smaller than Eq.~(\ref{eq: eimp}) and is proportional to $\Lambda^2$. 

%Let us focus first on the case where impurities are sparse enough that $n_i w^2 \ll 1$; i.e., where a single electron wave packet is unlikely to overlap with more than one impurity. We comment on the case $n_i w^2 \gg 1$ below.
As in the main text, we consider a disorder potential made of randomly distributed delta function impurities with strength $\pm \Lambda$. Let us imagine that the impurity potential (and therefore the positions of individual impurities) has some finite correlation length $\xi$ (which could reflect, for example, the finite size of impurities).
Consider now a domain of some large size $L$. Each electron within the domain has an expected number $\sim n_i w^2$ of impurities overlapping with its wave function so that the average number of impurities within the domain that overlap with an electron is $N_{imp}^L \sim (L^2/R^2) n_i w^2$. If $N_{imp} \gg 1$, then there is a statistical fluctuation in the number of such electrons of order $\sqrt{N_{imp}^L}$ from one domain to another. This fluctuation produces a corresponding fluctuation in the energy per electron, averaged across the domain:
\be
\delta \varepsilon_{imp}^L \sim \frac{\Lambda}{w^2} \frac{\sqrt{\left(L^2/R^2\right) n_i w^2}}{\left(L^2/R^2\right)}.
\ee

The WC can take advantage of these fluctuations by distorting through transverse shear, such that regions of size $L$ experience a shear displacement $c \sim \max\left\{ \xi, w \right\}$. Such a displacement effectively allows the region of size $L$ to sample a statistically independent area occupied by electron wave packets. 
The energy associated with shearing the region of size $L$ is $\mu c^2$, where $\mu \sim e^2/R^3$ is the shear modulus \cite{ruzin1992pinning}. The corresponding shear energy per electron is
\be
\varepsilon_{def} \sim \frac{\mu c^2}{\left(L^2/R^2\right)}.
\ee
 Minimizing $- \delta \varepsilon_{imp}+\varepsilon_{def}$, we arrive at the optimum domain size $L$:
\be
L \sim \frac{w e^2}{\Lambda \sqrt{n_i}} \frac{c^2}{R^2}.
\ee
This optimal domain size $L$ can be thought of as the correlation length of the Wigner lattice. (The small displacements $\sim r_0$ of individual electrons that are discussed in the main text do not disrupt long-ranged order since they are independent of each other.) The resulting energy per particle is given by:
% \begin{align}
%     \begin{split}
%         \varepsilon_{L}&\sim - \frac{n_{i}\Lambda^{2}R^3}{e^{2}w^{2}c^{2}},\\
%         & \sim - \frac{e^{2}}{a_{B}}\left(\frac{\Lambda}{e^{2}a_{B}}\right)^{2}\left(n_{i}a_{B}^{2}\right)\frac{1}{\left(R/a_{B}\right)^{1/2}}\left(\frac{R}{c}\right)^2.
%     \end{split}
%     \label{eq: eshear}
% \end{align}
\begin{align}
    \begin{split}
        \varepsilon_{L}&\sim-\frac{n_{i}\Lambda^{2}R^{3}}{e^{2}w^{2}c^{2}},\\&\sim-n_{i}\frac{\Lambda^{2}m}{\hbar^{2}}\frac{w^{2}}{c^{2}}.
    \end{split}
     \label{eq: eshear}
\end{align}
Since $c \sim \max\left\{ \xi, w \right\}$ and we have assumed $n_i^{-1/2} \gg w \sim (a_B R^3)^{1/4}$, it follows that the energy reduction per electron due to shear is no larger than the result of Eq.~(\ref{eq: eimp}) arising from independent electron displacements. If the impurity positions are completely uncorrelated, $\xi \rightarrow 0$, the result of Eq.~(\ref{eq: eshear}) is equivalent to Eq.~(\ref{eq: eimp}) up to an overall numerical coefficient, and if $\xi \gg w$ then it is parametrically smaller.

\end{document}